\documentclass[11pt]{article}
\usepackage{amsmath}
\usepackage{amsfonts}
\usepackage{amssymb}
\usepackage[dvips]{epsfig}

 \advance \textwidth by -2in
 \advance \textheight by -3in
\def\##1{{\underline #1}}
\def\~#1{{\underline {\mathcal#1}}}
\def\+#1{{{\mathcal #1}}}
\def\=#1{\underline{\underline #1}}

\def\.{\mbox{ \tiny{$^\bullet$} }}

\def\eps{\epsilon}
\def\epso{\epsilon_o}

\def\un{\#u_n}
\def\ub{\#u_b}
\def\utau{\#u_\tau}
\def\ux{\#u_x}
\def\uy{\#u_y}
\def\uz{\#u_z}

\def\muo{\mu_o}
\def\lambdao{\lambda_o}
\def\etao{\eta_o}
\def\ko{k_o}
\def\co{c_o}

\def\vkap{\varkappa}

\begin{document}

\noindent{\bf MORPHOLOGICAL EFFECTS ON SURFACE PLASMON POLARITONS AT THE PLANAR INTERFACE OF A METAL AND
A COLUMNAR THIN FILM}

\vskip 18 pt

\noindent \textbf{John A. Polo Jr$^1$ and Akhlesh Lakhtakia$^2$}\\
\vskip0.1cm
\noindent $^1$Department of Physics and Technology,\\
\noindent Edinboro University of Pennsylvania,\\
\noindent Edinboro, PA 16444, USA.\\
\noindent e--mail: polo@edinboro.edu\\
\noindent fax:  (814) 732-2455\\
\vskip 0.1cm
\noindent $^2$CATMAS --- Computational \& Theoretical Materials Sciences Group,\\
\noindent Department of Engineering Science and Mechanics,\\
\noindent  Pennsylvania State University,\\
\noindent University Park, PA 16802, USA.\\
\noindent e--mail: akhlesh@psu.edu

\vskip 18pt

\noindent {\bf ABSTRACT} Surface plasmon polaritons (SPPs) at the interface of a columnar thin film
(CTF) and metal exist over a range of propagation directions relative to the morphology of the CTF which
depends on the tilt of the columns in the CTF. The phase speed of the SPP wave varies mainly as a
function of the tilt of the CTF columns.  Both the confinement of the SPP wave to the interface and the
decay of the SPP wave along the direction of propagation depend strongly on the direction of propagation
relative to the morphologically significant plane of the CTF.  The greater the columnar tilt in relation
to the interface, the shorter is the range of propagation. Because of its porosity and the ability to
engineer this biaxial dielectric material, the CTF--metal interface may be more attractive than
traditional methods of producing SPPs.   \\

\vskip 6pt

\noindent {\bf Keywords:} {\it  surface plasmon polariton, columnar thin film, titanium oxide}

\section*{1.  INTRODUCTION}
\label{sect:intro}  % \label{} allows reference to this section
The surface plasmon polariton (SPP) has been the object of intense investigation for several decades
\cite{Pitarke}. In recent years, the understanding of the phenomenon has been harnessed for a wide array
of analytical and biomedical applications \cite{HYG99,Homola03,ACM06}. Although various configurations
are used to launch and detect the SPP, in the Kretschmann configuration, a metal film is interposed
between a low-refractive-index dielectric medium and a high-refractive-index dielectric medium \cite{KR68}. A light beam is
launched in the high-refractive-index medium to impinge on the planar interface between that medium and
the metal. The SPP travels along the planar interface of the metal and the low-refractive-index medium.

The propagation of SPPs at the planar interface of a metal and an anisotropic dielectric material  has
been examined theoretically \cite{AgMills}. For this communication, we investigated the propagation of
the SPP along the interface of a metal and a columnar thin film (CTF), in relation to the morphology of
the CTF. The columnar thin film, an artificial material, is effectively an optically biaxial continuum
\cite{HodgWuBook}, and porous \cite[Ch. 6]{STFbook}. The porous CTF may offer a medium in which to embed
analyte and/or recognition molecules to which the analyte may bind.

\begin{figure}
\begin{center}
\includegraphics[height=5.5cm]{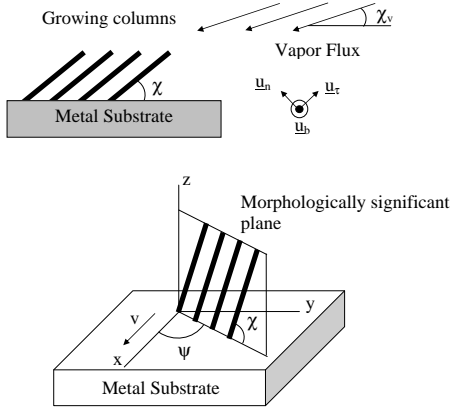}
\end{center}
\caption{Geometry of CTF--metal interface and SPP--wave propagation.}\label{Fig:Geometry}
\end{figure}

A columnar thin film is produced by directing a vapor at an angle $\chi_v$ to a substrate.  Under
suitable conditions, parallel columns form at an angle $\chi\geq\chi_v$ to the substrate; see Figure
\ref{Fig:Geometry}.  The columns are composed of multimolecular clusters with $\sim 3$ nm diameter
which, in turn, form columns with $\sim 100$ nm cross--sectional diameter, depending on  the evaporant
species and the deposition conditions. The geometry considered here is the same as that in a predecessor
paper~\cite{CTF06}, except that the substrate  is now made of a metal instead of a transparent
dielectric, and is also illustrated in Figure \ref{Fig:Geometry}.

The plan of the communication is as follows.  Section 2 contains a description of the geometry of the problem and the
constitutive relations. For the details of the field representation and the derivation of the dispersion
relation for the SPP wave, we refer the reader to the predecessor paper \cite{CTF06}.  Numerical results for a particular CTF--metal interface,
titanium oxide CTF--aluminum, are presented in Section 3.  Concluding remarks are presented in
Section 4.

The following assumption and notational conventions are used. An
$\exp(-i\omega t)$ time--dependence is implicit, with $\omega$ denoting the angular frequency. The
free--space wavenumber, the free--space wavelength, and the intrinsic impedance of free space are
denoted by $\ko=\omega\sqrt{\epso\muo}=\omega/\co$, $\lambdao=2\pi/\ko$, and $\etao=\sqrt{\muo/\epso}$,
respectively, with $\muo$ and $\epso$ being  the permeability and permittivity of free space. Vectors
are underlined, and dyadics are underlined twice. Cartesian unit vectors are
identified as $\ux$, $\uy$ and $\uz$.

\section*{2.  GEOMETRY AND CONSTITUTIVE RELATIONS}
The geometry of the problem is illustrated in Figure \ref{Fig:Geometry}.  With the interface between the
metal substrate and the CTF denoted by $z=0$, the metal
occupies the half--space $z<0$, while the CTF
occupies the half--space $z>0$.  Without loss of generality, the direction of propagation is taken
to be parallel to the $x$--axis, and the angle between the direction of propagation and the
morphologically significant plane \cite[Ch. 7]{STFbook} of the the CTF is denoted by $\psi$ .

The relative permittivity dyadic of the CTF may be stated as \cite[Ch. 7]{STFbook}
\begin{equation}
\=\eps_{CTF}=\left(n_a^2\,\un\un+n_b^2\,\utau\utau+n_c^2\,\ub\ub\right)\,,\label{Eqn:epsilon}
\end{equation}
where $n_{a,b,c}$ are the principal refractive indexes and the unit vectors
\begin{equation}
\left.\begin{array}{l}
\un= -(\ux \cos\psi+\uy \sin\psi)\sin\chi+\uz\cos\chi\\
\utau= (\ux \cos\psi+\uy \sin\psi)\cos\chi+\uz\sin\chi\\
\ub=-\uy\cos\psi+\ux\sin\psi
\end{array}\right\}\,.
\end{equation}
All four quantities~---~$n_{a,b,c}$ and the column inclination angle $\chi$~---~depend on the evaporant
species and the vapor incidence angle $\chi_v$. The refractive index of the metal is denoted by
$n_s$.

Expressions for the field equations and dispersion relation have been worked out elsewhere \cite{CTF06}. It suffices to mention that the wave vector in the substrate may be written as
\begin{equation}
\#k_s=\ko\,(\vkap\, \ux- iq_s\, \uz)\,, \label{kSub}
\end{equation}
where $\vkap$ and
\begin{equation}
q_s=+\sqrt{\vkap^2-n_m^2}\,\label{eqn:dispSub}
\end{equation}
are the normalized propagation constant and the decay constant, respectively.  In order for the SPP wave
to be confined to the interface $z=0$, we must have ${\rm Re}\left[q_s\right]>0$.  Unlike for the
CTF--dielectric interface \cite{CTF06}, $\vkap$ is expected to be complex valued since the dissipative
properties of the metal can not be ignored. Similarly, in the half--space $z>0$ occupied by the CTF, the
wavevector may be written as
\begin{equation}
\#k_c=\ko\,( \vkap\, \ux+ iq_c\, \uz)\,, \label{kCTF}
\end{equation}
and again we must have ${\rm Re}\left[q_c\right]>0$ for localization of the SPP wave to the interface.
The SPP wave comprises two partial waves in the CTF, identified by $q_{c1}$ and $q_{c2}$.

\section*{3.  NUMERICAL RESULTS}
Hodgkinson {\em et al.} \cite{Hodgkinson} measured the constitutive relations for CTFs made of oxides of
tantalum, titanium, and zirconium.  We chose titanium oxide to illustrate SPP propagation at a
planar CTF--metal interface. The empirical relationships  determined at
$\lambda_o=633$ nm for the titanium--oxide CTF can be written as
\begin{eqnarray}
\label{eqH1}
n_a&=&1.0443+2.7394\left(\frac{\chi_v}{\pi/2}\right)-1.3697\left(\frac{\chi_v}{\pi/2}\right)^2 \,,\\
n_b&=&1.6765+1.5649\left(\frac{\chi_v}{\pi/2}\right)-0.7825\left(\frac{\chi_v}{\pi/2}\right)^2 \,,\\
n_c&=&1.3586+2.1109\left(\frac{\chi_v}{\pi/2}\right)-1.0554\left(\frac{\chi_v}{\pi/2}\right)^2 \,,
\end{eqnarray}
and
\begin{eqnarray}
\tan\chi&=&2.8818\tan\chi_v\,, \label{eqH4}
\end{eqnarray}
where $\chi_v$ and $\chi$ are in radian.  We must caution that the foregoing expressions are applicable
to CTFs produced by one particular experimental apparatus, but may have to be modified for  CTFs
produced by other researchers on different apparatuses; hence, we used these expressions for the numerical results
presented in this section merely for illustration.  Metals commonly used for SPP systems are aluminum,
copper, silver, and gold.  For   illustration, we chose aluminum with a complex refractive index
$n_s=1.38 + 7.61 i$ at $\lambda_o=633$ nm~\cite{Mansuripure}. We note that values of $\vkap$ for SPP propagation when $\psi=0^\circ$ have been reported elsewhere \cite{LPajp}.

\begin{figure}
\begin{center}
\begin{tabular}{c}
\includegraphics[height=5.5cm]{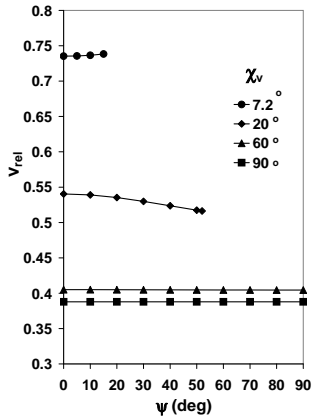}
\end{tabular}
\end{center}
\caption{Phase speed of the SPP wave at the planar interface of a titanium--oxide CTF and an aluminum substrate, relative to
the speed of light in  vacuum, versus $\psi$ for $\chi_v=7.2^\circ$, $20^\circ$, $60^\circ$, and
$90^\circ$.}\label{Fig:vrel}
\end{figure}

Calculations were carried out for $\chi_v=7.2^\circ, 20^\circ, 60^\circ$, and $90^\circ$.  The lowest
value $\chi_v=7.2^\circ$ corresponds to $\chi=20^\circ$ and represents the approximate, currently
achievable lower limit of $\chi$.  As previously stated, the wave number $\kappa=\ko\vkap$ along the direction of propagation
must be complex valued.  The real part of $\kappa$ determines the phase speed $v=\omega/{\rm Re}(\kappa)$ of the SPP
wave.  The phase speed relative to the speed of light in  vacuum is calculated as
\begin{equation}
v_{rel}=\frac{v}{\co}=\frac{1}{{\rm Re}(\vkap)}\,.\label{Eq:Speed}
\end{equation}
 Figure \ref{Fig:vrel} shows $v_{rel}$ as a
function of $\psi$ for $\chi_v=7.2^\circ, 20^\circ, 60^\circ$, and $90^\circ$, with $\psi$ restricted to
the range $[0^\circ,90^\circ]$.
The curves are drawn over the $\psi$--ranges for which the boundary conditions at $z=0$ could be satisfied and
thus represent the ranges over which  a SPP wave can exist. For each value of $\psi$ shown, at which a
SPP wave exists, there also exist three other values at $-\psi$, $180^\circ+\psi$, and
$180^\circ-\psi$ with an identical value of $v_{rel}$. The relative phase speed
$v_{rel}$ decreases as $\chi_v$
increases with the most rapid change occurring at the low end of the $\chi_v$--range.

The greatest variation of $v_{rel}$ as a function of $\psi$ is observed for $\chi_v=20^\circ$.  This
curve has a downward slope as $v_{rel}$ decreases from roughly 0.540 to 0.516 as $\psi$ changes from
$0^\circ$ to $52^\circ$.  At $\chi_v=60^\circ$, a much smaller change is observed:
 $v_{rel}$ decreases
from roughly 0.4050 to 0.4045 as $\psi$ changes from $0^\circ$ to $90^\circ$.  In contrast, the curve
for $\chi_v=7.2^\circ$ shows an upward slope as $v_{rel}$ changes from roughly 0.735 to 0.738 as $\psi$
varies from $0^\circ$ to $15^\circ$.  Of course, the curve for $\chi_v=90^\circ$ is completely flat as
required by symmetry since Eqs. (\ref{eqH1})--(\ref{eqH4}) predict that the generally biaxial CTF
becomes uniaxial at $\chi_v=90^\circ$ with the axis of symmetry perpendicular to the interface $z=0$.

\begin{figure}
\begin{center}
\begin{tabular}{c}
\includegraphics[height=5.5cm]{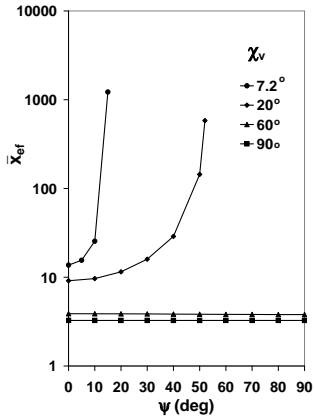}
\end{tabular}
\end{center}
\caption{e--folding distance of the SPP wave along the $x$--direction, the direction of propagation, versus $\psi$ for
the same conditions as in Figure \ref{Fig:vrel}.}\label{Fig:efolding_x}
\end{figure}

The imaginary part of $\kappa$ determines the rate at which the SPP wave decays along the direction of
propagation.  The e--folding distance for decay of the wave amplitude relative to the wavelength
$\lambda_o$ may be calculated as
\begin{equation}
\overline{x}_{ef}=\frac{1}{2\pi \,{\rm Im}(\vkap)}\,.\label{Eq:efolding_x}
\end{equation}
Figure \ref{Fig:efolding_x} shows $\overline{x}_{ef}$ as a
function of $\psi$ over the restricted range
$[0^\circ,90^\circ]$.  As with $v_{rel}$, for each value of $\psi$ displayed, an identical value of
$\overline{x}_{ef}$ is obtained at $-\psi$, $180^\circ +\psi$, and $180^\circ -\psi$.  As $\chi_v$ increases,
$\overline{x}_{ef}$ decreases and the curve of $\overline{x}_{ef}$ vs. $\psi$ flattens.  At $\chi_v=60^\circ$, the curve is nearly flat
with only $\sim2\%$ change over the entire  $\psi$--range. In contrast, at $\chi_v=7.2^\circ$ and
$20^\circ$, $\overline{x}_{ef}$ changes by nearly two orders of magnitude.

\begin{figure}
\begin{center}
\begin{tabular}{ccc}
\includegraphics[height=5.5cm]{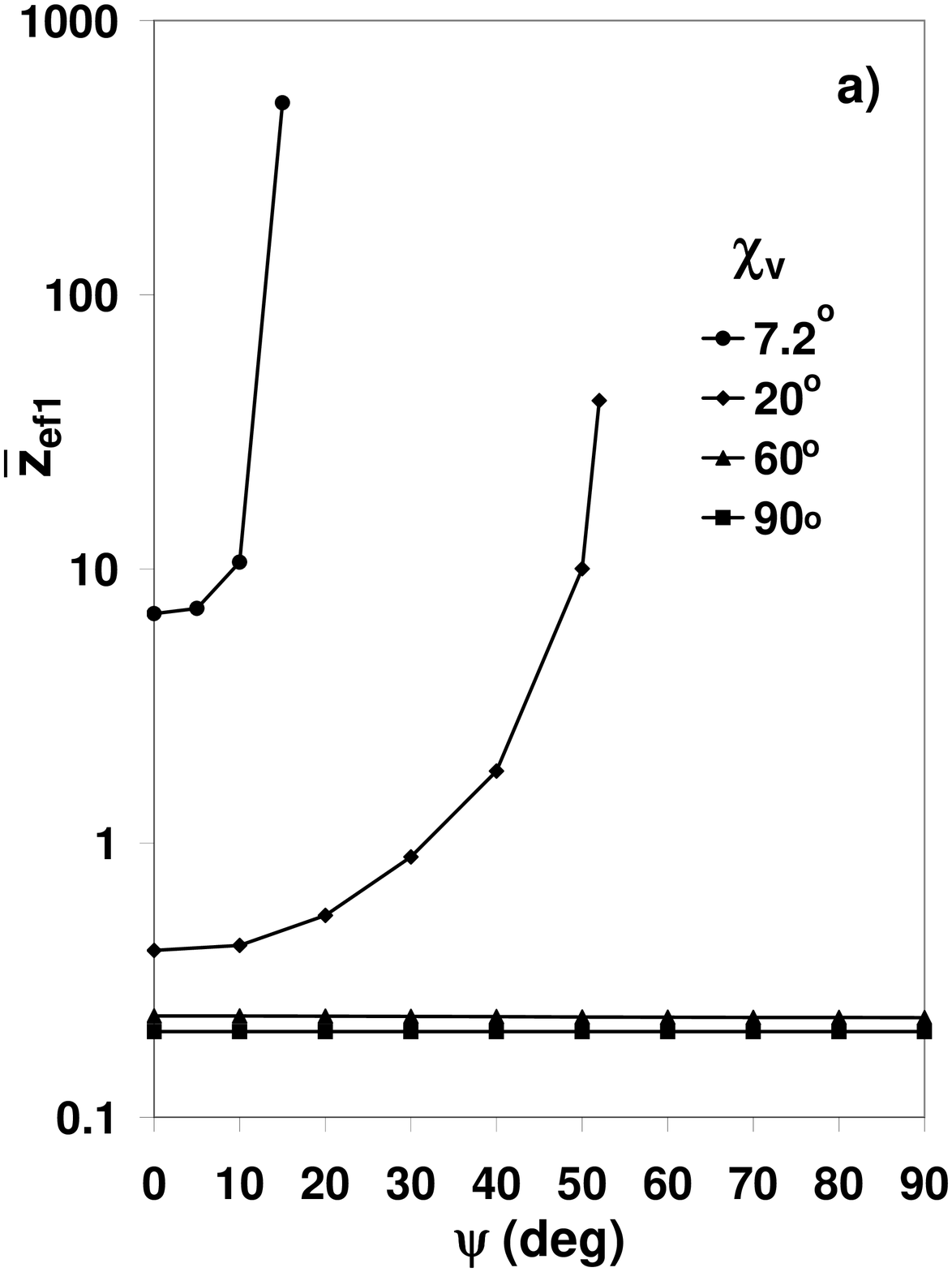}
\includegraphics[height=5.5cm]{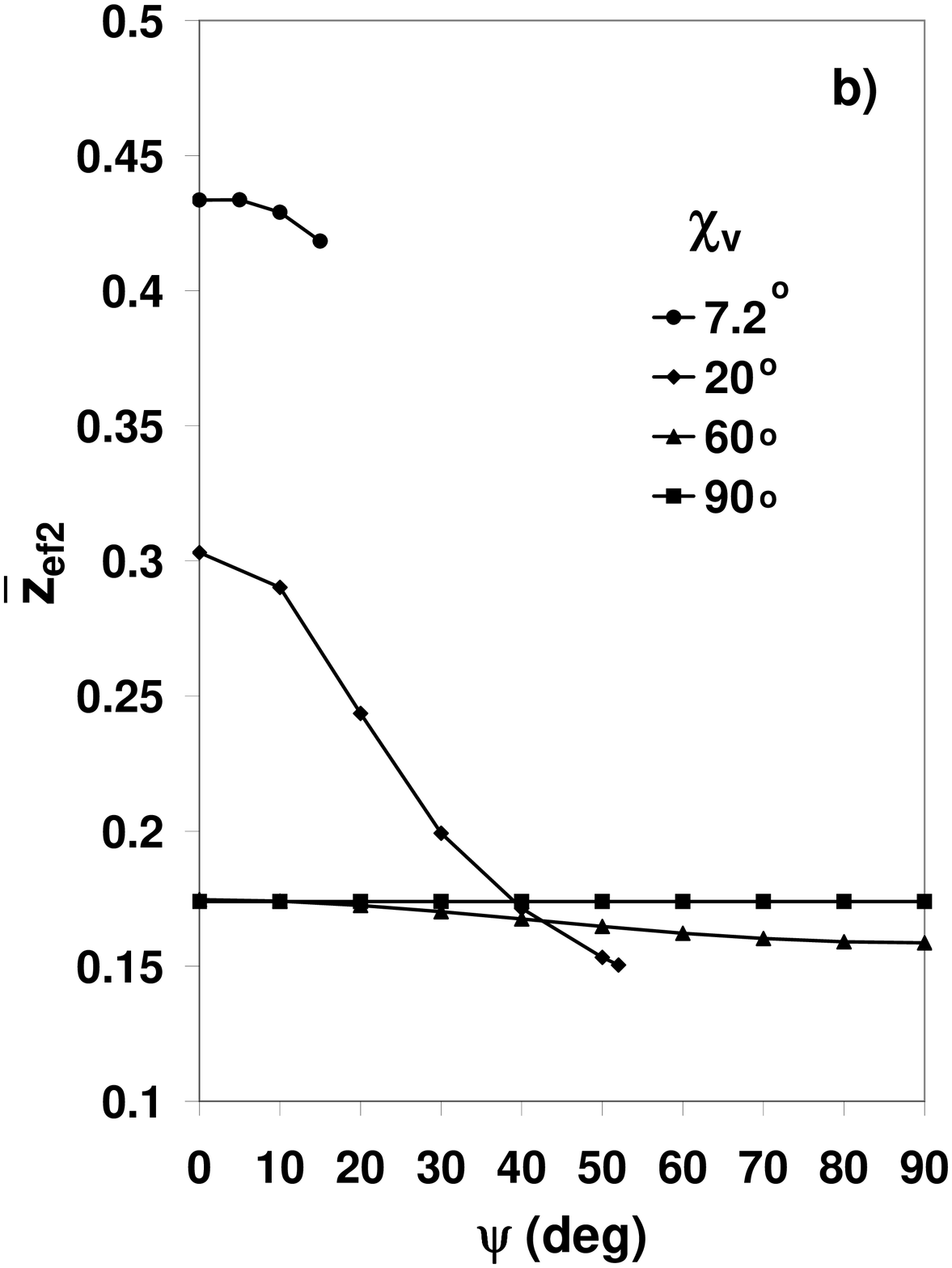}
\includegraphics[height=5.5cm]{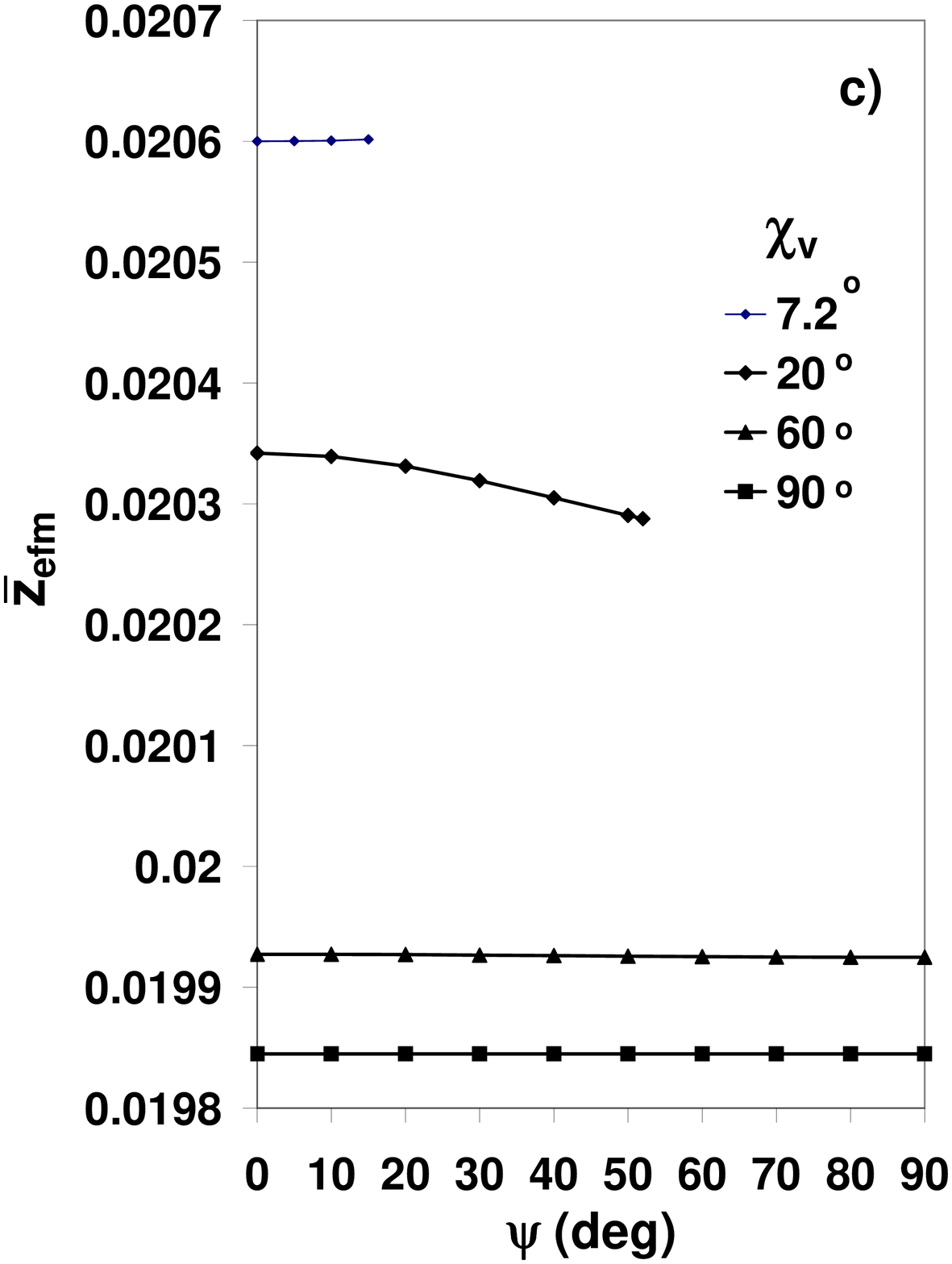}
\end{tabular}
\caption{e--folding distances of the SPP wave relative to $\lambda_o$ in the $z$--direction for the same
conditions as in Figure \ref{Fig:vrel}:  a)  partial wave 1 in CTF;  b)  partial wave 2 in CTF;  c)  in
aluminum.}\label{Fig:efolding_z}
\end{center}
\end{figure}

 Figures \ref{Fig:vrel} and  \ref{Fig:efolding_x} let us conclude
that an increase in the vapor incidence angle (i) reduces  the phase speed and (ii) increases the
attenuation of the SPP wave. Thus, a high value of $\chi_v$ is inimical to long--range propagation for
all $\psi$, an understanding previous obtained only for the case of the direction of propagation lying
in the morphologically significant plane (i.e., $\psi=0^\circ$) \cite{LPajp}. The properties of the SPP
wave along the $z$--axis, perpendicular to the direction of propagation, are described by
\begin{itemize}
\item[(i)] $q_{c1}$ and (ii) $q_{c2}$ for the two partial waves in the CTF, and
\item[(iii)] $q_m$ in the metal.
\end{itemize}
Similar to
$\overline{x}_{ef}$, three e--folding distances in the $z$--direction relative to the wavelength in vacuum
may be calculated as
\begin{equation}\left.
\begin{array}{lll}
\overline{z}_{ef1}&=&\frac{1}{2\pi\, {\rm Im}(q_{c1})}\\[5pt]
\overline{z}_{ef2}&=&\frac{1}{2\pi\,{\rm  Im}(q_{c2})}\\[5pt]
\overline{z}_{efm}&=&\frac{1}{2\pi\, {\rm Im}(q_{m})}
\end{array}\right\}\,.
\end{equation}
Figure \ref{Fig:efolding_z} shows these three quantities as
functions of $\psi$ over the restricted range
$[0^\circ,90^\circ]$.
These three quantities  describe the localization of the SPP wave to the interface.  In Figure
\ref{Fig:efolding_z}a, the curves of $\overline{z}_{ef1}$ are similar to those of $\overline{x}_{ef}$
displayed in Figure \ref{Fig:efolding_x}, with the curves for both $\chi_v=60^\circ$ and $90^\circ$
essentially flat.  At $\chi_v=7.2^\circ$ and $20^\circ$, $\overline{z}_{ef1}$ varies by about two
orders of magnitude.  Thus, for the two values of $\chi_v$ for which the $\psi$--range for surface wave
propagation ends and does not continue to $\psi\geq 90^\circ$, the first partial--wave component of the SPP wave in the CTF becomes
delocalized as $\psi$ approaches the end of its range.

In Figure \ref{Fig:efolding_z}b, the curves of $\overline{z}_{ef2}$ versus $\psi$ are also nearly flat
for both $\chi_v=60^\circ$ and $90^\circ$.  At $\chi_v=7.2^\circ$ and $20^\circ$, $\overline{z}_{ef2}$
decreases as $\psi$ increases.  The second partial--wave component, then, becomes more localized as
$\psi$ approaches the end of its range.

The e--folding distance in the metal, $\overline{z}_{efm}$, is
essentially flat for all values of $\chi_v$  in Figure \ref{Fig:efolding_z}c.  The
greatest variation of $\overline{z}_{efm}$ occurs when $\chi_v=20^\circ$, but is still less than 0.4\%.

\begin{figure}
\begin{center}
\begin{tabular}{cc}
\includegraphics[height=5.5cm]{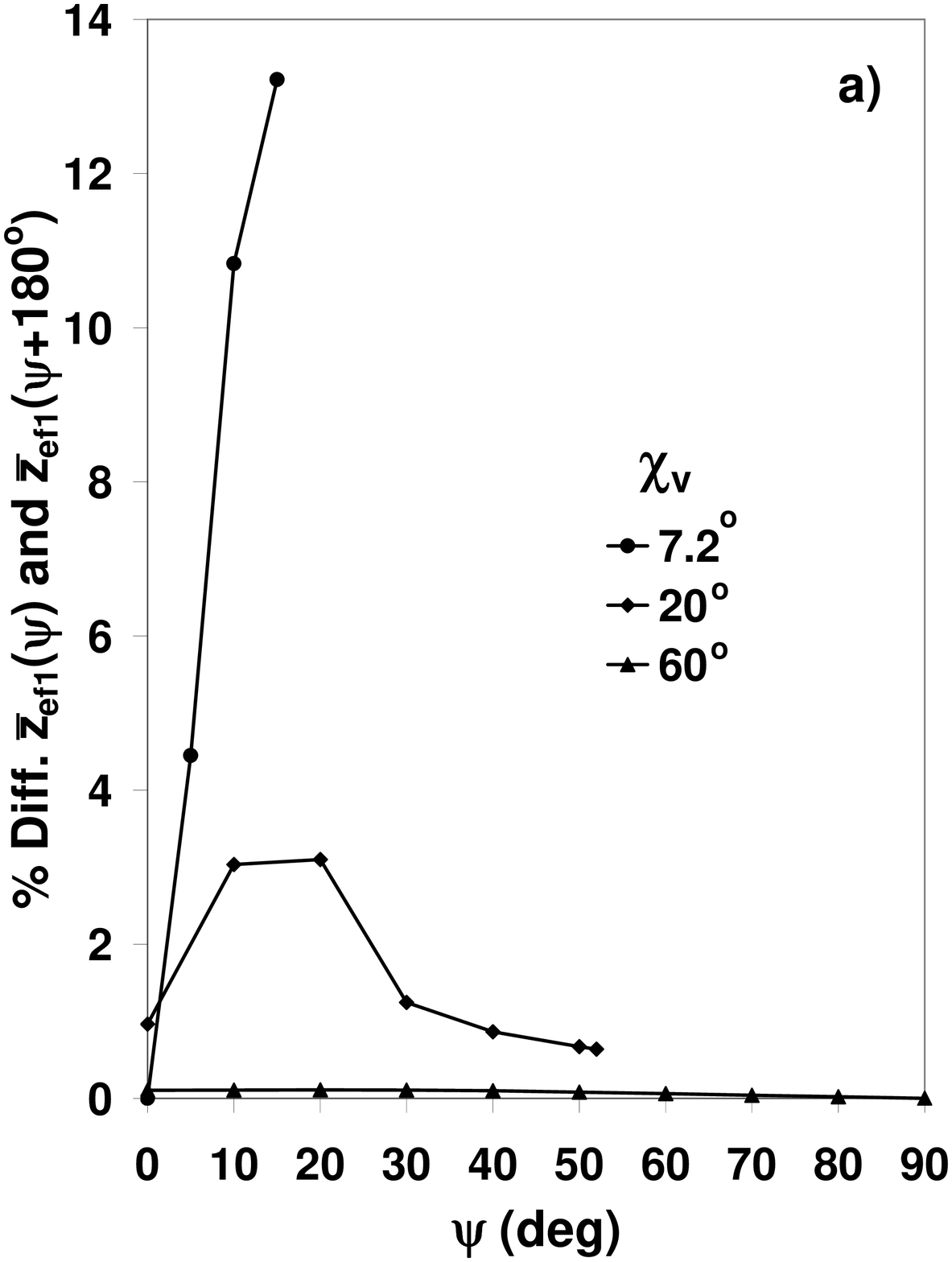}
\includegraphics[height=5.5cm]{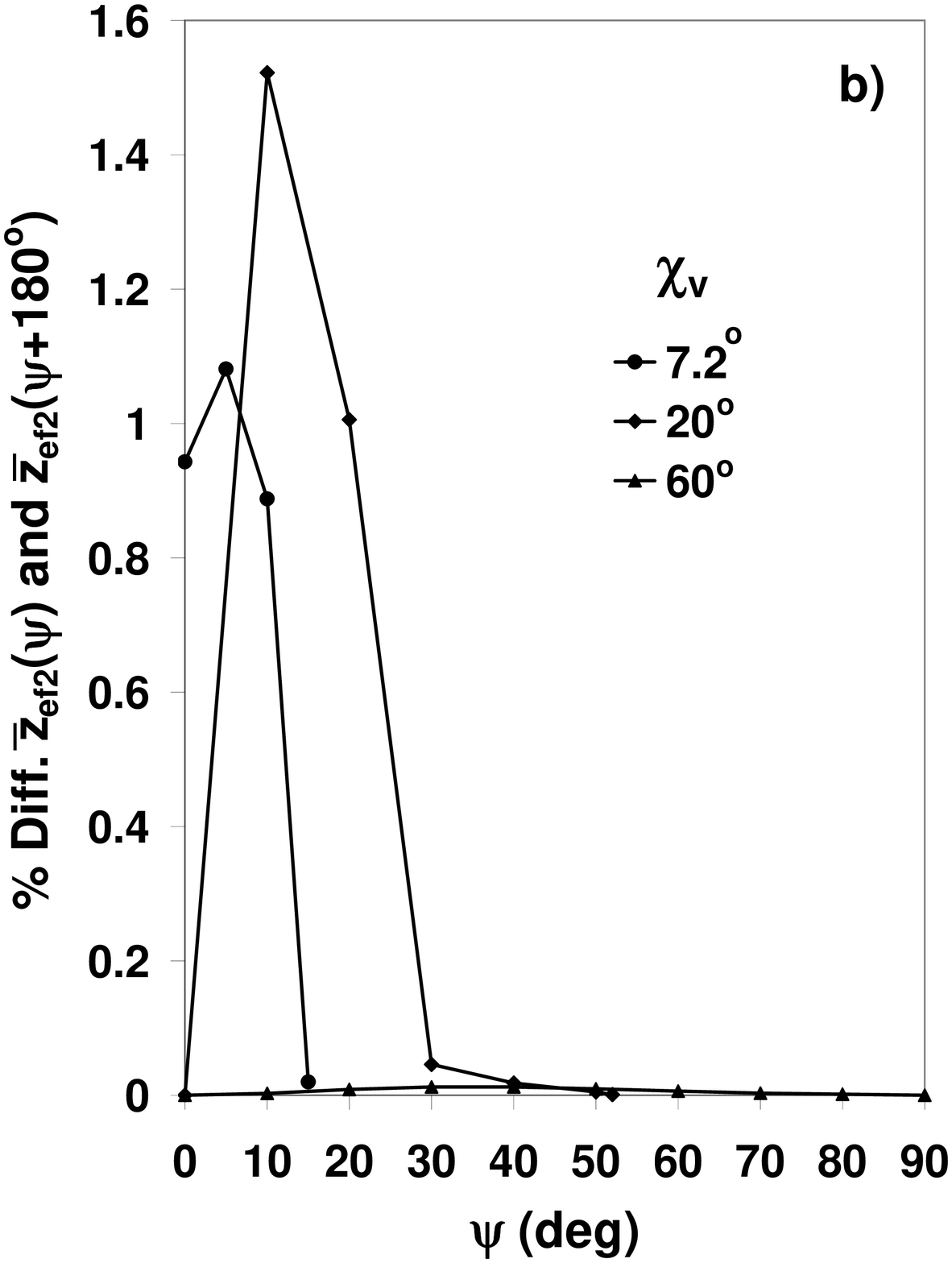}
\end{tabular}
\caption{The \% difference between $\overline{z}_{ef1}(\psi)$ and $\overline{z}_{ef2}(180^\circ+\psi)$
for same conditions as in Figure \ref{Fig:vrel}.}\label{Fig:efolding_diffs}
\end{center}
\end{figure}

Unlike $v_{rel}$ and $\overline{x}_{ef}$, the values of the e--folding distances in the $z$--direction
for the CTF are not identical at $\psi$, $-\psi$, $180^\circ+\psi$, and $180^\circ-\psi$.  The value of
any e--folding distance in the $z$--direction at $\psi$ is, however, equal to the value at $-\psi$, and
the value at $180^\circ+\psi$ is equal to that at $180^\circ-\psi$.  Since the difference in most cases
is slight, we present the percent difference between the values at $\psi$ and $-\psi$, and the values at
$180^\circ+\psi$ and $180^\circ-\psi$ in Figure \ref{Fig:efolding_diffs}.  The \% differences for the
e--folding distance of the first partial--wave component in the CTF, $\overline{z}_{ef1}$, are shown in
Figure \ref{Fig:efolding_diffs}a.  The curve for $\chi_v=7.2^\circ$ shows the largest difference of
about 13\% and occurs at the end of the $\psi$--range where $\psi=15^\circ$.  The difference decreases
rapidly with increasing $\chi_v$.  At $\chi_v=60^\circ$, the maximum difference is only about 0.1\%.
Differences in e--folding distances for the partial--wave component 2 in the CTF are shown in Figure
\ref{Fig:efolding_diffs}b with the maximum observed value of 1.5\%~---~ almost an order of magnitude
smaller than for the partial--wave component 1. Differences in e--folding distances appear to decrease
as $\chi_v$ increases with a maximum difference at $\chi_v=60^\circ$ of only about 0.01\%. However, the
decrease is not monotonic as the maximum difference at $\chi_v=20^\circ$ is about 50\% larger than that
at $\chi_v=7.2^\circ$.

\section*{4.  CONCLUSION}
Examining SPP--wave propagation at a planar CTF--metal interface, with demonstrated empirical data to
characterize both the CTF and the metal, we found that the direction, denoted by $\psi$, at which SPP--wave propagation is possible is limited, in general, and depends on the tilt angle  $\chi$ of the
columns in the CTF. This tilt angle may be predicted from the vapor incidence angle $\chi_v$ set during
the manufacture of the film. The $\psi$--range for SPP waves increases as $\chi_v$ increases. At
sufficiently large values of $\chi_v$, SPP propagation becomes possible in all directions.  Even at the
lowest value $(\chi_v=7.2^\circ$) examined by us, the $\psi$--range is greater than $10^\circ$.  This is
in contrast to surface--wave propagation at the interface of a CTF and an isotropic dielectric material, for which the $\psi$--range  is only a fraction of a degree for all values of $\chi$ \cite{CTF06}.

The phase speed of the SPP wave shows a strong dependence on $\chi_v$, but is also mildly dependent on
$\psi$.  The e--folding distance along the direction of propagation is strongly dependent on $\psi$ at
low values of $\chi_v$, varying by several orders of magnitude, but is relatively flat at larger values
of $\chi_v$. A high value of $\chi_v$ is inimical to long--range propagation for all $\psi$. At the end
of the $\psi$--range, for those values of $\chi_v$ for which the propagation directions are limited, the
SPP wave becomes delocalized from the interface on the CTF side.  This is seen as an apparent divergence
in the e--folding distance in the direction perpendicular to the interface for one of the two partial
waves in the CTF as $\psi$ approaches the end of its range.  The e--folding distance perpendicular to
the interface for the other partial wave in the CTF, on the other hand, tends to decrease as $\psi$
approaches its limiting value.  In the metal, the e--folding distance perpendicular to the interface
does not vary much with $\psi$.

The demonstration of a SPP wave at the interface of a CTF and a metal offers certain options in SPP
technology.  With a wide choice of possible evaporant materials~\cite{Macleod} and vapor incidence angles, the
CTF--metal interface offers a large latitude in the engineering of SPP systems.  The porosity of the CTFs
may also offer some advantages.  With its nano--scale porous structure, a CTF could allow molecular--scale analytes to reach the interface for analysis while screening out larger particulates.  The
detection of viruses has become important in recent times.  CTFs with pores on the nano--scale may allow
viruses access to the interface while excluding larger organisms.  Through photolithographic techniques
it is possible to pattern CTFs~\cite{Horn04}.  Various samples could be placed, possibly in the field, on
isolated  CTF patches on a single substrate for later analysis.  Also, different patches could be embedded with
different recognition molecules which could be used to perform multiple searches on a single sample.
Other options have yet to be imagined.


\begin{thebibliography}{99}
\bibitem{Pitarke}
J.~M. Pitarke, V.~M. Silkin, E.~V. Chulkov, and P.~M. Echenique, Theory of surface plasmons and
surface-plasmon polaritons, Rep. Prog. Phys. 70 (2007), 1-87.
%


\bibitem{HYG99}
J. Homola, S.~S. Yee, G. Gauglitz, Surface plasmon resonance sensors: review, Sens. Actuat. B: Chem. 54
(1999), 3-15.
%
\bibitem{Homola03}
J. Homola, Present and future of surface plasmon resonance biosensors, {Anal. Bioanal. Chem.} {377},
528-539 (2003).
%
\bibitem{ACM06}
S.~K. Arya, A. Chaubey, and B.~D. Malhotra,
Fundamentals and applications of biosensors, {Proc. Indian Natn. Sci. Acad.}
{72} (2006), 249-266.
%
\bibitem{KR68}
E. Kretschmann and H. Raether, Radiative decay of nonradiative surface plasmons excited by light, {Zeit. Naturforsch. A} 23 (1968), 2135-2136.
%
\bibitem{AgMills}
V.~M. Agranovich and D.~L. Mills (eds.), Surface polaritons: Electromagnetic waves at surfaces and
interfaces, North-Holland, Amsterdam, 1982.
%
\bibitem{HodgWuBook}
I.~J. Hodgkinson  and Q.~h. Wu, Birefringent thin films and polarizing elements, World Scientific,
Singapore, 1997.
%
\bibitem{STFbook} A. Lakhtakia and R. Messier, {Sculptured thin
films: Nanoengineered morphology and optics}, SPIE Press, Bellingham, WA, 2005.
%
\bibitem{CTF06}
J.~A. Polo, Jr., S.~R. Nelatury, and A. Lakhtakia, Propagation of surface waves at the planar interface
of a columnar thin film and an isotropic substrate, J. Nanophoton. {1} (2007), 013501.
%
\bibitem{Hodgkinson} I.~J. Hodgkinson, Q.~h. Wu, and J. Hazel,
Empirical equations for the principal refractive indices and column angle of obliquely deposited films
of tantalum oxide, titanium oxide, and zirconium oxide, {Appl. Opt.} {37} (1998), 2653-2659.
%
\bibitem{Mansuripure}M. Mansuripur and L. Li, What in the world are surface
plasmons?,
{OSA Opt. Photon. News\/}, 8 (1997) 50-55 (May issue).

\bibitem{LPajp} A. Lakhtakia and J.~A. Polo, Jr.,
Morphological influence on surface--wave propagation at the planar interface of a metal film and a
columnar thin film, Asian J. Phys. (accepted for publication, 2007); also: arXiv:0706.4306.
%
\bibitem{Macleod}
H.~A. Macleod, Thin--film optical filters, 3$^{rd}$ ed., CRC Press, Boca Raton, FL, 2001.
%
\bibitem{Horn04}
M.~W. Horn, M.~D. Pickett, R. Messier, and A. Lakhtakia, Blending of nanoscale and microscale in uniform
large--area sculpured thin--film architectures, Nanotechnology 15 (2004), 303-310.

\end{thebibliography}
\end{document}